\documentstyle[12pt]{article}

\begin{document}

\baselineskip=7mm

\newcommand{\TeV}{\,{\rm TeV}}
\newcommand{\GeV}{\,{\rm GeV}}
\newcommand{\MeV}{\,{\rm MeV}}
\newcommand{\keV}{\,{\rm keV}}
\newcommand{\eV}{\,{\rm eV}}
\newcommand{\Tr}{{\rm Tr}\!}
\renewcommand{\arraystretch}{1.3}
\newcommand{\be}{\begin{equation}}
\newcommand{\ee}{\end{equation}}
\newcommand{\bea}{\begin{eqnarray}}
\newcommand{\eea}{\end{eqnarray}}
\newcommand{\ba}{\begin{array}}
\newcommand{\ea}{\end{array}}
\newcommand{\bmat}{\left(\ba}
\newcommand{\emat}{\ea\right)}
\newcommand{\refs}[1]{(\ref{#1})}
\newcommand{\ler}{\stackrel{\scriptstyle <}{\scriptstyle\sim}}
\newcommand{\ger}{\stackrel{\scriptstyle >}{\scriptstyle\sim}}
\newcommand{\lag}{\langle}
\newcommand{\rag}{\rangle}
\newcommand{\ns}{\normalsize}
\newcommand{\cm}{{\cal M}}
\newcommand{\gr}{m_{3/2}}
\newcommand{\p}{\partial}

\def\321{$SU(3)\times SU(2)\times U(1)$}
\def\tl{{\tilde{l}}}
\def\tL{{\tilde{L}}}
\def\bd{{\overline{d}}}
\def\tL{{\tilde{L}}}
\def\a{\alpha}
\def\b{\beta}
\def\g{\gamma}
\def\c{\chi}
\def\d{\delta}
\def\D{\Delta}
\def\db{{\overline{\delta}}}
\def\Db{{\overline{\Delta}}}
\def\e{\epsilon}
\def\l{\lambda}
\def\n{\nu}
\def\m{\mu}
\def\nt{{\tilde{\nu}}}
\def\p{\phi}
\def\P{\Phi}
\def\x{\xi}
\def\r{\rho}
\def\s{\sigma}
\def\t{\tau}
\def\th{\theta}
\def\ne{\nu_e}
\def\nm{\nu_{\mu}}

\begin{titlepage}
\title{ {\large\bf LIGHT SINGLET FERMIONS AND NEUTRINO PHYSICS}\\
                                          \vspace{-3cm}
                                          \hfill{\ns IC/95/240\\}
                                          \hfill{\ns hep-ph/9508329\\}
                                          \vspace{3cm} }

\author{Eung Jin Chun$^1$\thanks{Talk presented at the International
              Workshop on Elementary Particle Physics:
              Present and Future, Valencia 95.} \hspace{.4cm}
         Anjan S.~Joshipura$^2$ \hspace{.4cm}
         Alexei Yu.~Smirnov$^{1,3}$ \\[.5cm]
  {\ns\it $^1$International Center for Theoretical Physics}\\
  {\ns\it P.~O.~Box 586, 34100 Trieste, Italy} \\[.3cm]
  {\ns\it $^2$Theoretical Physics Group, Physical Research Laboratory}\\
  {\ns\it Navarangpura, Ahmedabad, 380 009, India} \\[.3cm]
  {\ns\it $^3$Institute for Nuclear Research, Russian Academy of Sciences}\\
  {\ns\it 117312 Moscow, Russia} }
\date{}
\maketitle
\begin{abstract} \baselineskip=6.5mm
{\ns  The existence of a light singlet fermion mixed with the electron
neutrino is hinted by the simultaneous explanation of various neutrino
anomalies.  We show that supersymmetry can provide a natural framework for
the existence and the desired properties of such a fermion.
Quasi Goldstone fermions (QGF) of spontaneously broken global symmetries
like the Peccei-Quinn symmetry or lepton number can mix properly with the
neutrinos provided the presence of the $R$-parity breaking term $\e LH_2$.
The lightness of QGF can be a consequence of non-minimal
K\"ahler potentials like that of no-scale supergravity.
In order to keep $R$-parity, such a sterile component has to
be placed in a new singlet superfield with no vacuum expectation value.
In the context of the standard seesaw mechanism the lightness of such a
singlet can be understood by imposing a $R$-symmetry.
}\end{abstract}
\thispagestyle{empty}
\end{titlepage}

\section{Introduction}

All the experimentally known fermions transform non-trivially
under the gauge group \321 of the standard model (SM).
However there are  experimental
hints in the neutrino sector which suggest the existence of \321
- singlet fermions mixing appreciably with the known neutrinos.
These hints come from
(a) the deficits in the solar \cite{solar} and atmospheric \cite{atm}
neutrino fluxes
(b) possible need of significant hot component \cite{dm} in the dark
matter of the universe  and
(c) some indication of $\bar{\nu}_e-\bar{\nu}_{\mu}$ oscillations in
the laboratory \cite{lsnd}.
These hints can be reconciled with each
other if there exists a fourth very light ($< {\cal O}$(eV))
neutrino mixed with some of the known  neutrinos preferably
with the electron one.  The fourth neutrino is
required to be sterile in view of the strong bounds on number of
neutrino flavours coming both from the LEP experiment
and from the primordial nucleosynthesis \cite{ns}.

The existence of a very light sterile neutrino demands theoretical
justification since unlike the active neutrinos, the mass of a
sterile state is not protected by the gauge symmetry of the SM
and hence could be very large.  Usually a sterile neutrino is
considered on the same footing as the active neutrinos and some
ad hoc symmetry is introduced to keep this neutrino light.
Recently there are several attempts to construct models for sterile
neutrinos which have their origin  beyond the usual lepton
structure \cite{paper1,paper2,mirror1,mirror2,ma}.

In this report, we discuss the role of supersymmetry (SUSY) in explaining
both the existence and the lightness of a singlet fermion $S$ which can mix
with the neutrinos.  As a case of special interest we will concentrate
on the mass of $S$ and its mixing with the electron neutrino in the range:
\bea \label{parameters}
 m_S &\simeq& (2-3)\cdot 10^{-3} \eV \nonumber\\
 \sin\th_{es} &\simeq& \tan\th_{es} \simeq (2-6)\cdot 10^{-2} \;.
\eea
These values of parameters allow one to solve the solar neutrino
problem through the resonance conversion $\n_e \to S$ \cite{msw}.
More discussions on simultaneous reconciliations of the diverse neutrino
problems can be found in refs.~\cite{paper1,paper2} on which this report is
based.

\section{Quasi Goldstone Fermion}

The existence of SM-singlet fields is a common property in physics
beyond the standard model.
The most interesting examples are the Goldstone bosons of spontaneously
broken global symmetries required to solve the strong CP problem (the
Peccei-Quinn symmetry) \cite{pq} and to explain the origin of neutrino masses
(the lepton number symmetry) \cite{cmp}.
In the SUSY limit, a spontaneously broken global
symmetry automatically generates a massless singlet (Goldstone) fermion
being a superpartner of a Goldstone boson.  However, SUSY breakdown
results in generation of mass of a Goldstone fermion.
While the existence of these quasi Goldstone fermions (QGF)
is logically independent of neutrino physics, there are good
reasons to expect that these fermions will couple to neutrinos.
Indeed, in the case of lepton number symmetry the superfield which is
mainly responsible for the breakdown of the lepton number symmetry
carries nontrivial lepton number and therefore it can directly couple
to leptons if the charge is appropriate.
In the case of the PQ symmetry,
this superfield could couple to the Higgs supermultiplet.
If theory contains small violation of $R$-parity then this mixing
with the Higgs gets communicated to the neutrino sector.
Thus the occurrence of a QGF can have implications for neutrino
physics.
In the following subsections we elaborate upon  the expected properties
of the QGF: their masses arising after SUSY breaking and
the mixing of these fermions with the electron neutrino.

\subsection{masses of QGF}

The supersymmetric standard model with some global symmetry $U(1)_G$ can be
characterized by the following superpotential:
 \be \label{w}
 W=W_{MSSM}+W_S+W_{mixing} \;,  \ee
where $W$ is assumed to be invariant under $U(1)_G$.
As we outlined in the above, this symmetry may be identified with the
PQ symmetry, lepton number symmetry or combination thereof. The
first term in eq.~\refs{w} refers to the superpotential of the minimal
supersymmetric standard model (MSSM).
The second term contains \321 singlet superfields which are responsible
for the breakdown of $U(1)_G$.
The minimal choice for $W_S$ is
\be \label{ws} W_S=\lambda (\s \s'- f_G^2) y \;, \ee
where $\s,\s'$ carry non trivial $G$-charges and $f_G$ sets the
scale of $U(1)_G$ breaking.
The last term of eq.~\refs{w} describes mixing of the singlet
fields with the superfields of the MSSM.

In the case \refs{ws} the Goldstone fermion is  contained in
$S\sim \s-\s'$ and is massless in the SUSY limit.
Broken SUSY itself cannot automatically protect the mass of a QGF.
It depends on the structure of the superpotential
$W_S$~\cite{chun1} and  on the pattern of soft-terms~\cite{chun2}.
It also depends on the way this breaking is communicated to the singlet $S$
and the scale $f_G$ \cite{paper2}.
The most natural framework for light QGF is no-scale
supergravity~\cite{noscale}.  No-scale models
contain only one kind of soft-terms, namely, gaugino masses.
Therefore, th soft SUSY-breaking terms corresponding to $W_S$ in eq.~\refs{w}
are absent at tree-level and thus QGF remains massless.
However, the radiative mass can be triggered by the \321 gaugino masses
through a set of interactions.
A realistic example can be found in the context of the seesaw mechanism.
The vacuum expectation value (VEV) of the field
$\s$ (or $\s'$) may give rise to large masses of right-handed (RH)
neutrinos $N$ as in the following superpotential invariant under
$U(1)_G$:
 \be\label{seesaw}
 W = {m^D \over \lag H_2 \rag} L N H_2 + \frac{M}{f_{G}} N N \s \;,  \ee
where we have omitted the generation indices.
The generation structure of the superpotential \refs{seesaw} will depend on
the $U(1)_G$-charge assignment to the fields \cite{paper2}.
This $U(1)_G$ symmetry is not necessarily the lepton number symmetry as we
will discuss in subsection 2.2.
The first term in eq.~\refs{seesaw} gives rise to the Dirac masses
of the neutrinos, whereas the second one gives the Majorana masses of
RH neutrino components.
The scale $f_{G} \sim 10^{10} - 10^{12} \GeV$ generates
$M \sim 10^{10}-10^{11}$ GeV required by the hot dark matter
and atmospheric neutrinos.
If the soft-term $A_N NN\s$ with $A_N \sim m_{3/2}$ is present, there appears
one-loop mass of the QGF proportional to $A_N$ \cite{ckl}.
But in no-scale models $A_N=0$ at tree-level and the QGF mass is
indeed generated in three loops as shown in Figure~1.
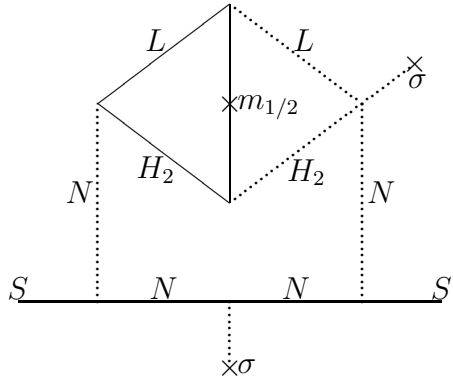
\begin{figure}
\begin{picture}(200,150)(-220,-20)
\put(-80,0){\line(160,0){160}}
\multiput(0,0)(0,-2.5){10}{\circle*{.1}}
\multiput(-50,0)(0,2.5){30}{\circle*{.1}}
\multiput(50,0)(0,2.5){30}{\circle*{.1}}
\put(-50,75){\line(4,3){50}}
\put(-50,75){\line(4,-3){50}}
\multiput(50,75)(-2,1.5){25}{\circle*{.1}}
\multiput(50,75)(-2,-1.5){25}{\circle*{.1}}
\multiput(50,75)(2,1.5){10}{\circle*{.1}}
\put(0,37.5){\line(0,75){75}}
\put(0,75){\makebox(0,0){$\times$}} \put(3,73){$m_{1/2}$}
\put(70,90){\makebox(0,0){$\times$}} \put(67.5,81){$\sigma$}
\put(0,-25){\makebox(0,0){$\times$}} \put(3,-27){$\sigma$}
\put(-80,5){\makebox(0,0){$S$}}
\put(80,5){\makebox(0,0){$S$}}
\put(-25,5){\makebox(0,0){$N$}}
\put(25,5){\makebox(0,0){$N$}}
\put(-28,99){\makebox(0,0){$L$}}
\put(28,99){\makebox(0,0){$L$}}
\put(-28,50){\makebox(0,0){$H_2$}}
\put(29,49){\makebox(0,0){$H_2$}}
\put(-62,37.5){$N$}
\put(52,37.5){$N$}
\end{picture}
\caption{Three-loop diagram for the QGF mass.
 The cross with $m_{1/2}$ denotes gaugino mass insertion.}
\end{figure}
This three-loop mass can be  estimated as
 \be \label{radmass3}
 m_S \simeq {\a_2\over (4\pi)^5} {m_{\n} M^3 \over v_2^2 f_{G}^2} m_{1/2}
 \;. \ee
Here $\a_2$ and  $m_{1/2}$ are the $SU(2)$ fine structure constant and
gaugino mass respectively.
For $m_\n \simeq 3$ eV, $m_{1/2} \simeq v_2 \simeq 100$ GeV, and $f_G \simeq
10^{12}$ GeV, one gets $m_S \simeq 3 \cdot 10^{-3}$ eV with a value of
$M \simeq 10^{10}$ GeV.

A contribution to the mass of the QGF can follow also from
interactions, $W_{mixing}$, which mix $S$ with usual neutrinos
(subsection 2.2).

\subsection{Neutrino-QGF mixing}

We now discuss how the QGF can mix with neutrinos.
Such a mixing implies the violation of $R$-parity
conventionally imposed in the MSSM \cite{hall}. This is simply because that
the leptons being ordinary matter fields are $R$-even and the QGF being a
fermionic partner of a Goldstone boson is $R$-odd.
The violation of $R$-parity may destabilize the lightest supersymmetric
particle (LSP) which is usually considered as the cold dark matter (CDM)
of the Universe.
For this reason,  we consider  the PQ symmetry as a good candidate for
$U(1)_G$ since the coherent oscillation of the axion can provide the CDM
for $f_{PQ} \sim 10^{12}$ GeV \cite{kt}.
Therefore, the PQ mechanism required for a resolution of the strong CP
problem can supply both the CDM and the sterile neutrino.

The best way to implement the PQ symmetry in the MSSM is to extend the Higgs
mass term in such a way that the smallness of the Higgs mass parameter $\mu$
can be naturally obtained.  For instance, let us consider the
non-renormalizable term \cite{mu}
 \be \label{nr}
 \lambda H_1H_2\frac{\s^2}{M_P}\;, \ee
where $M_P$ is the Planck mass \footnote{One can also introduce the
renormalizable term to generate $\mu \simeq m_{3/2}$ \cite{chun3}.}.
Here the VEV of $\s$, $\lag \s \rag
\sim f_{PQ}$, spontaneously breaks the PQ symmetry.
In this case, $\mu= \lambda\frac{\lag\s\rag^2}{M_P}$ can be about the weak
scale.  When the axion superfield $S$ is predominantly consists of $\s$,
the PQ symmetry breaking yields the Higgs mass term and the coupling of $S$
to the Higgs superfields
\be\label{mix1} W_{mixing}=c_\m\frac{\mu}{f_{PQ}}H_1H_2S + \m H_1H_2 \ee
with $c_\m$ being ${\cal O}(1)$.
In order to have the mixing of $S$ with neutrinos, one needs the lepton
number violating term $\e LH_2$.
It is remarkable  to notice  that the PQ scale is in the right range for
the RH neutrino masses.
The PQ symmetry can indeed play a role of the lepton number symmetry if
both the Higgs and leptons transform non-trivially under the PQ symmetry
as in ref.~\cite{lpy}.  In this case one can correlate
the origin of $\e$ and $\m$ to the same symmetry breaking scale $f_{PQ}$.
The neutrino and Higgs coupling to QGF is then given by
 \bea\label{mix3}
 W_{mixing}&=&\m H_1H_2+\epsilon L_eH_2 + \nonumber\\
 & & c_{\m}\frac{\mu}{f_{PQ}}H_1H_2S +
 c_{\e}\frac{\e}{f_{PQ}}L_eH_2S \;, \eea
where $L_e$ is the electron doublet.
If the PQ symmetry is the standard one unrelated to
the lepton sector, the parameter $\e$ vanishes.
On the other hand, the global $U(1)$ symmetry becomes
the usual lepton number symmetry when $c_\m = 0$ and the bare $\mu$-term is
introduced.

An example of models which leads to the mixing terms of eq.~\refs{mix3}
can be obtained  by the PQ-charge prescription ($-1$,$-1$, 1,$-1$,$-2$) for
($H_1$, $H_2$, $\s$, $\s'$, $L_e$). It  permits
the following $U(1)_{PQ}$ invariant superpotential:
 \be \label{model2}
 W =  \l (\s\s' - f_{PQ}^2)y + {\d_\m \over M_P} H_1 H_2 \s^2
 +{\d_\e \over M_P^2} L_e H_2 \s^3 \;,  \ee
which gives the terms displayed in eq.~(\ref{mix3}) with
$c_\e={3\over\sqrt{2}},c_\m=\sqrt{2}$.
\smallskip

The $W_{mixing}$ in eq.~\refs{mix3}
generates the following effective mass matrix for $\n_e$ and $S$
 \be\label{matrix3}
 \left( \ba{cc}
 0&(c_\e-c_\m) \e v\sin\beta/ f_{PQ}\\
 (c_\e-c_\m) \e v\sin\beta/ f_{PQ}
 &m_S^0- c_{\m}^2\m v^2 \sin2\beta / f_{PQ}^2\\
 \ea \right) \;, \ee
where we added the direct mass $m_S^0$ which can be generated by the
mechanism of subsection 2.1.
According to eq.~\refs{matrix3} the $\n_e-S$ mixing angle $\theta_{es}$
is determined by
 \be\label{ts2}
 \tan \theta_{es}\sim \frac{(c_\m-c_\e) \e v\sin\beta}
 {m_S^0 f_{PQ}- c_{\m}^2\m v^2 \sin2\beta / f_{PQ}} \;. \ee
For $f_{PQ} \simeq 10^{12}$ GeV, $m_S^0 \simeq 3\cdot 10^{-3}$ eV is the
dominant contribution to the mass of $S$. In this case one obtains from
eq.~(\ref{ts2}) for the  $\nu_e-S$ mixing
 \be \label{ts1}
 \tan \theta_{es}\sim \frac{\e v \sin \beta}
 {m_S^0 f_{PQ}}\;.  \ee
Then the desired value, $\tan \theta_{es} \sim (2 - 6)\cdot 10^{-2} \eV$
\refs{parameters}, can be obtained if the $R$-parity breaking parameter
$\e$  equals
 \be \label {epsis}
 \e \sim \frac{m_S^0 f_{PQ} \tan \theta_{es}}{v \sin\beta}
 \approx (2 - 6)\cdot 10^{-16} \frac{f_{PQ}}{\sin \beta} \;.  \ee
For $f_{PQ} \sim 10^{12}$ GeV one has $\e \sim 0.1$ MeV.

Let us remark the other possibilities for the QGF mass.
If $m_S^0 = 0$ in eq.~\refs{matrix3}, the QGF mass,
$m_S = (2 - 3)\cdot 10^{-3} \eV$
can be obtained for  the marginally allowed value of the PQ scale:
 \be f_{PQ} \approx v \sqrt{\frac{\mu \sin 2\beta}{m_S}}
 \ler 4 \cdot10^9 \GeV \;.  \ee
For $f_{PQ} > 10^{10}$ GeV the QGF mass generated via $\mu$-term
is too small for the MSW solution. For $f_{PQ} \sim 10^{11}$ GeV,
$m_S  \approx 10^{-5} \eV$ is in the region of ``just-so" solution
of the solar neutrino problem.
In these cases, however, axions cannot provide the CDM as we noted before.

\section{A light singlet in the standard seesaw structure}

In the previous case, the QGF mixes with the electron neutrino directly ($\e
c_\e \neq 0$) or via its coupling to the Higgses ($c_\m \neq 0$).  The small
mass of the QGF was related to the multi-loop effect or the suppression by
$1/f_{PQ}^2$ due to the Goldstone property. An important consequence was the
$R$-parity violation leading to destabilization of the LSP.

In this section, we will suggest another scheme in which $R$-parity is
preserved.  For this, one should place the singlet $S$ in the superfield with
zero VEV. This implies that the singlet has to be introduced from outside.
Being a singlet $S$ can mix with neutrinos via its coupling to the
right-handed neutrinos.  In this case, the existence of $S$ cannot be
explained but the smallness of its mass can be understood in terms of
the seesaw mechanism.  In order to implement a light singlet fermion in the
standard seesaw structure, we will suggest to use $R$-symmetry which
occurs in many SUSY theories.  The (unbroken) $R$-parity is then
embedded in the $R$-symmetry.
\smallskip

Let us first determine the parameters appearing in the phenomenological
superpotential
 \be \label{base}
 W = {m_e \over \lag H_2 \rag}L_e N_eH_2 + {M_e\over 2}N_eN_e
              + m_{es}N_e S \;, \ee
where $N_e$ is the right-handed neutrino component.
The Dirac mass $m_e$ and the mixing mass $m_{es}$
are much smaller than the Majorana mass $M_e$: $m_e, m_{es} \ll M_e$.
The superpotential \refs{base} leads to the mass matrix in the basis
$(S, \n_e, N_e)$:
\be \label{mm1}
 {\cal M} = \bmat{ccc} 0 & 0 & m_{es}\\ 0 & 0 & m_e \\ m_{es} & m_e & M_e
              \emat \;.
\ee
The diagonalization of \refs{mm1} is straightforward. One combination of
$\n_e$ and $S$ is massless and the orthogonal combination
acquires a mass via the see-saw mechanism:
\be \label{m1}
 m_1 \simeq -{m_e^2 + m_{es}^2 \over M_e}\;.
\ee
The mass of the heavy neutrino is $\simeq M_e$.
The $\n_e$--$S$ mixing angle is determined by
\be \label{th}
 \tan\th_{es} = {m_e \over m_{es}} \;.
\ee
Taking for $m_e$ the typical Dirac mass of the first generation:
$m_e \sim (1-5) \MeV$, and suggesting that $\n_e \rightarrow
S$ conversion explains the solar neutrino problem with $m_1 =m_S$
as in \refs{parameters}, we find
\be \label{mis}
 m_{es} = {m_e \over \tan\th_{es}} \simeq (0.02-0.3) \GeV \;.
\ee
According to \refs{m1} the RH mass scale is
\be \label{Mis}
 M_e \simeq m_{es}^2/m_1 = {m^2_e \over m_1 \tan^2\th_{es}}
   \simeq (10^8-3\cdot10^{10})\GeV\;.
\ee
One has now to understand how the mixing mass \refs{mis} arises without
introducing new mass scales.
One also has to ensure that there is no direct coupling of $S$ with $L_e$,
and the mass term $SS$ is absent or negligibly small.
\smallskip

Our prescription is quite simple.  Consider the superpotential
\be \label{model3}
 W = {m_e\over \lag H_2\rag} L_e N_eH_2 + f N_e N_e\s + f' N_e S y
        - {\l \over 2}(\s^2 - M^2)y \;.  \ee
whose structure is determined by the $R$-symmetry
under which the fields ($L_e$, $N_e$, $S$, $y$, $\s$, $H_2$) carry
the $R$-charges (1, 1,$-1$, 2, 0, 0).
Note that the $R$-symmetry forbids the bare mass terms $SS$ as well as
the coupling $SS\s$.  The last term in eq.~\refs{model3} can be replaced by
$(\s\s'-M^2)y$ to implement the lepton number symmetry.
In the global SUSY limit, $\s$ gets non-zero
VEV $\lag \s \rag \simeq M \sim 10^{11}$ GeV which generates the Majorana
mass of $N_e$: $M_{e} = f \lag \s \rag$.
The point is that $y$ develops a VEV as a consequence of SUSY breaking.
Broken SUSY produces the following soft-breaking terms in the scalar
potential:
 \bea \label{soft}
 V_{soft} &=&  \{ A_L {m_e \over \lag H_2 \rag} L_e N_eH_2 +
 fA_\n  N_e N_e\s + f' A_S N_eSy - \nonumber\\
 & & {\l \over2}(A_y\s^2 - B_yM^2)y + \mbox{h.c.} \} +
 \sum_i m_i^2 |z_i|^2 \;, \eea
where $z_i$ denotes the fields appearing in the superpotential \refs{model3}
and $A_L$, etc., are the soft-breaking parameters.
Minimization of the potential shows the following:
(1) The fields $L_e, N_e, S$ do not develop VEV and therefore $R$-parity is
unbroken.
(2) The field $y$ acquires non-zero VEV due to the soft-breaking terms.
Consequently, the mixing mass for $S$ and $N_e$ appears:
 \be  \label{mis2}
 m_{es}= {f'\over 2\l}(A_y-B_y) \ee
Since $m_{es} \gg m_1$, no strong tunning of $A_y-B_y$ is needed.
For $A_y-B_y \sim {\cal O}(m_{3/2})$,
the desired value of $m_{es}$ \refs{mis}
can be obtained by choosing $f'/\l \sim 10^{-3}-10^{-2}$.
However, more elegant possibility is that $A_y=B_y$ at tree level
but a non-zero value  for $A_y - B_y$ is generated due to radiative
corrections through the differences in interactions of $\s$ and $y$.
In this case  one expects
\be \label{mrad}
  m_{es} \sim {\bar{\l}^2\over 16\pi^2} m_{3/2} \;,
\ee
where $\bar{\l}$ represents a combination of the constants $\l,f$ and $f'$.
As a consequence, the value $m_{es}\sim 0.1$ GeV does not require
smallness of $\bar{\l}$ or $f'$.

The equality $A_y = B_y$ at tree level can be achieved  by the
introduction of non-minimal K\"ahler potential allowing
mixings between the observable and hidden sectors.
Let us introduce the following K\"ahler potential:
\be
 K = C\overline{C} + C\overline{C}(a\frac{Z}{M_{Pl}} +
     \overline{a}\frac{\overline{Z}}{M_{Pl}}) + Z\overline{Z} \;,
\ee
where $C$ and $Z$ represent an observable and hidden sector field,
respectively.  Then usual assumption that the observable sector has no
direct coupling to the hidden sector in superpotential, $W=W(C) + W(Z)$,
leads to the universal soft-terms:
\be
 V_{soft} \sim m_{3/2} W(C) +  \mbox{h.c.} \;,
\ee
provided $\overline{a}= \lag W(Z) \rag / \lag M_{Pl} \partial W/\partial Z +
W(Z)\overline{Z}/M_{Pl} \rag $.
Note also that the field $C$ does not acquire a soft-breaking mass.
This mechanism can be generalized to arbitrary number of observable sector
fileds.  For our purpose  $C \equiv \s, y$, i.e., we couple $\s$ and $y$ to
the hidden sector field $Z$ with the above-mentioned choice for $a$.

\section{Conclusions}

Simultaneous presence of different neutrino anomalies points to the
existence of a sterile neutrino.  In particular, the resonance conversion
of the electron neutrino into such a singlet fermion $S$ can explain
the solar neutrino problem provided its mass and mixing are appropriate
\refs{parameters}.
Supersymmetry is shown to provide a framework within which the existence
and the desired properties of such a light fermion follow naturally.

We have considered first a possibility that the sterile neutrino is
a quasi Goldstone fermion appearing in supersymmetric theories
as a result of spontaneous breaking of a global $U(1)_G$ symmetry.
This global $U(1)_G$ symmetry can be identified with the PQ
symmetry, the lepton number symmetry.
The smallness of $m_S$ can be attributed in supergravity theory
to no-scale kinetic terms for certain superfields.
The mixing of QGF with the neutrinos implies spontaneous or explicit
violation of $R$-parity.  QGF can mix with neutrino via interaction with
Higgs multiplets (in the case of PQ symmetry) or directly via coupling with
the combination $L H_2$ (in the case of lepton number symmetry).
In the case of the PQ symmetry, the PQ-scale $f_{PQ}\sim 10^{10}-10^{12}$ GeV
determines several features of the model presented here.
It provides simultaneous explanation of the parameters $\e$ and $\m$ and
thus leads to small $R$-parity  violation ($\e LH_2$ with $\e \sim 0.1$ MeV)
required in order to solve the solar neutrino problem in our approach.
It also provides the  intermediate scale
for the right-handed neutrino masses which is required in order to solve
the dark matter and the atmospheric neutrino problem. Furthermore,
it controls the magnitude of the radiatively generated mass of the QGF
and allows it to be in the range needed for the
MSW solution of the solar neutrino problem.
Finally, the CDM can consist of the axion if $f_{PQ} \sim 10^{12}$ GeV.
Thus the basic scenario presented here is able to correlate variety of
phenomena.

The conservation of R-parity requires for the fermion $S$ to be
a component of singlet superfield which has no VEV.
This allows to construct simple model \refs{model3}  in which the
properties (mass and mixing) of $S$ follow from the conservation of
$R$-symmetry.  The singlet field is mixed with RH neutrinos by the
interaction with the field $y$ which can acquire VEV radiatively after soft
SUSY breaking.

Let us finally comment on the other phenomenological consequences of the
existence of such a sterile state $S$.
An $U(1)_G$ symmetry being generation-dependent \cite{paper1,paper2}
can provide simultaneous explanations for the predominant coupling of $S$ to
the first generation (thus satisfying the nucleosynthesis bound)
and for the pseudo-Dirac structure of $\n_\m$--$\n_\t$ needed in solving
the atmospheric neutrino and the hot dark matter problem.
In this case, it appears nontrivial to accommodate the
parameters of $\bar{\n}_\m \to \bar{\n}_e$ oscillations in the region of
sensitivity of LSND and KARMEN experiments.
The simplest way is to introduce a slight violation of the $U(1)_G$ symmetry
through which such parameters can be incorporated.


\end{document}